\documentclass[12pt,letterpaper]{article}

\usepackage{amsmath,amssymb,calc}
\usepackage{graphicx}
\usepackage{cite}

\newcommand\be{\begin{equation}}
\newcommand\ee{\end{equation}}
\newcommand{\bea}{\begin{eqnarray}}
\newcommand{\eea}{\end{eqnarray}}

\newcommand{\nn}{\nonumber}
\newcommand{\pd}{\partial}

\def\id{\protect{{1 \kern-.28em {\rm l}}}}

\def\id{\protect{{1 \kern-.28em {\rm l}}}}

\begin{document}

\begin{titlepage}
\begin{center}
\hfill \\
\vspace{1cm}
{\Large {\bf Dark Energy from Inspiraling\\ in Field Space\\[3mm] }}

\vskip 1.5cm
{\bf Lilia Anguelova${}^a$\footnote{anguelova@inrne.bas.bg}, John Dumancic${}^b$\footnote{dumancjp@mail.uc.edu}, Richard Gass${}^b$\footnote{gassrg@ucmail.uc.edu}, L.C.R. Wijewardhana${}^b$\footnote{rohana.wijewardhana@gmail.com}\\
\vskip 0.5cm  {\it ${}^a$ Institute for Nuclear Research and Nuclear Energy,
Bulgarian Academy of Sciences, Sofia 1784, Bulgaria\\
${}^b$ Department of Physics, University of Cincinnati, Cincinnati, OH 45221, USA}}

\vskip 6mm

\end{center}

\vskip .1in
\vspace{0.5cm}

\begin{center} {\bf Abstract}\end{center}

\vspace{-1cm}

\begin{quotation}\noindent

We find an exact solution of the equations of motion of a two-field cosmological model, which realizes multi-field dark energy. The latter is characterized by field-space trajectories with turning rates that are always large. We study a class of two-field models and show that it is possible to have such trajectories, giving accelerated space-time expansion, even when the scalar potential preserves the rotational invariance of the field-space metric. For the case of Poincar\'e-disk field space, we derive the form of the scalar potential compatible with such background solutions and, furthermore, we find the exact solutions analytically. Their field-space trajectories are spirals inward, toward the center of the Poincar\'e disk. Interestingly, the functional form of the relevant scalar potential is compatible with a certain hidden symmetry, although the latter is broken by the presence of a constant term.

\end{quotation}

\end{titlepage}

\eject

\tableofcontents

\section{Introduction}

Understanding the nature of dark energy is one of the most important problems of modern cosmology. A standard explanation for the present-day accelerated expansion of the Universe is the presence of a cosmological constant. An appealing alternative, however, is that dark energy is dynamical, resulting from the evolution of one or more scalar fields. This possibility is more natural from the perspective of high energy physics, in view of the tiny value of the relevant energy density. Furthermore, recent conjectural conditions \cite{OOSV,GK,OPSV}, for effective field theories to be compatible with quantum gravity, strongly favor cosmological models with more than one scalar fields \cite{AP,BPR}.\footnote{The relevance of these conjectures for low energy physics is under a question mark though \cite{deAl}.}

Such multi-field models can lead to new effects, compared to single-field ones, when their background solutions have non-geodesic field-space trajectories. The latter are characterized by a non-vanishing turning rate function. Models with large turning rates have attracted a lot of attention in the literature on cosmological inflation. That Early Universe period is rather similar to the present-day accelerated expansion, despite the huge difference in energy scales. It is natural then to consider the possibility for multi-field dark energy models, relying on field-space trajectories with large turning rates.\footnote{It should be noted that the considerations of \cite{ACPRZ}, concerning rapid-turn inflation, are unlikely to be relevant for dark energy, because the supergravity structure of the effective action is not expected to be preserved at the extremely low-energy scales of interest for late-time cosmology. Nevertheless, we will comment more on \cite{ACPRZ} in Section \ref{Discussion}.} Such a proposal was explored in \cite{ASSV}, where it was argued that it can lead to distinguishing observational features, despite having equation-of-state parameter very close to $-1$\,.\footnote{For other dark energy models with multiple scalars see \cite{BCK,KLT,CFLR,JSM,DAHK,AKLV,CDP,CDP2,MSGW}.} However, it remained an open question whether there are actual cosmological solutions of this type for a well-defined scalar potential.

We will investigate a certain kind of two-field cosmological models and will find a class of exact solutions to their equations of motion, which realizes the proposal of \cite{ASSV}. Our focus will be the dark energy sector and we will comment on additional matter fields only briefly, when appropriate. As discussed in \cite{ASSV}, this captures the essential features of the proposed mechanism, which is natural since matter gets diluted by the expansion of the Universe, while dark energy (with equation-of-state parameter $\approx -1$) does not.

The two-field models we will study have rotationally invariant field-space metric and scalar potential. It was thought that this invariance is incompatible with expanding space-time solutions, whose field-space trajectories have constantly-large turning rates. By analyzing the background equations of motion, we show that, in fact, such solutions do exist. In the process, we derive a relation between the field space metric and the scalar potential, which has to be satisfied for solutions of this type. This relation can be viewed as an ODE determining the form of the potential, for a given choice of field-space metric. It simplifies considerably for hyperbolic field spaces with a fixed Gaussian curvature. Taking the field space to be the Poincar\'e disk, we then derive the scalar potential compatible with solutions of the desired type.

It turns out that the functional form of the potential is exactly what is required by the hidden symmetry of \cite{ABL}, although that symmetry is broken by an additive constant term. This enables us to find the background solution analytically by using techniques from the Noether symmetry method. The latter is a powerful tool for finding exact solutions, well-known from the context of extended theories of gravity; see, for instance, \cite{CR,CMRS,CNP,CDeF}. It was applied to two-field cosmological models with hyperbolic field spaces in \cite{ABL}. Adapting a result from that work allows us to achieve a great simplification of the equations of motion in the present case, thus enabling us to solve them analytically. 

The exact solutions we find have field-space trajectories, which are spiraling inward toward the center of the Poincar\'e disk. The corresponding Hubble parameter is always finite and tends fast to a constant, while the equation-of-state parameter approaches the value $-1$ arbitrarily closely. Notably, we show that the slow roll parameter is a monotonically decreasing function, which tends to zero, in the entire physical parameter-space of the solutions. It is worth underlining that, in our case, the phenomenologically-relevant part of field space is at small field values, unlike in the standard hyperbolic models in the literature that require very large ones.

This paper is organized as follows. In Section \ref{Two_f_Models_and_DE}, we review the equations of motion of two-field cosmological models with rotationally-invariant field spaces. We also explain the motivation 
behind the type of background solution we will be looking for, in order to model dark energy. 
In Section \ref{Sol_Generalities}, we analyze the equations of motion and derive the form of the scalar potential, which is compatible with the desired type of solutions for a Poincar\'e-disk field space.
In Section \ref{ExactSol}, we find analytically the exact solutions, following from this choice of field-space metric and the corresponding potential, by utilizing a close similarity to a certain hidden symmetry. In Section \ref{ParSpace}, we investigate the parameter space of our solutions, as well as the shape of their field-space trajectories. We also prove that the $\varepsilon$-parameter of the solutions is a monotonically decreasing function. Finally, in Section \ref{Discussion}, we summarize our results and discuss some implications, as well as directions for further research.

\section{Two-field cosmological models and dark energy} \label{Two_f_Models_and_DE}
\setcounter{equation}{0}

We will study a class of cosmological models arising from two scalar fields minimally coupled to Einstein gravity. Our goal will be to find an exact solution of the equations of motion of this system, suitable for a certain description of dark energy. In this Section, we begin by reviewing the relevant equations of motion and specializing them to the case of a rotationally-invariant scalar field space. We then outline the basic idea, which motivates the kind of dark energy model we are aiming to realize. 

\subsection{Equations of motion for rotationally-invariant field spaces}

The action for two scalar fields $\phi^I (x^{\mu})$ minimally coupled to gravity is the following:
\be \label{Action_gen}
S = \int d^4x \sqrt{-\det g} \left[ \frac{R}{2} - \frac{1}{2} G_{IJ} (\{\phi^I\}) \pd_{\mu} \phi^I \pd^{\mu} \phi^J - V (\{ \phi^I \}) \right] \,\,\, ,
\ee
where $g_{\mu \nu}$ is the spacetime metric with $\mu,\nu = 0,...,3$ and $G_{IJ}$ is the metric on the field space parametrized by the scalars $\{\phi^I\}$ with $I,J = 1,2$. We assume the standard cosmological Ansatze for the background spacetime metric and scalar fields, given by:
\be \label{metric_g}
ds^2_g = -dt^2 + a^2(t) d\vec{x}^2 \qquad , \qquad \phi^I = \phi^I_0 (t) \quad ,
\ee 
where $a(t)$ is the scale factor. As usual, the definition of the Hubble parameter is:
\be \label{Hp}
H (t) = \frac{\dot{a}}{a} \,\,\, , 
\ee
where we have denoted $\dot{} \equiv \pd_t$\,.

The equations of motion for $H(t)$ and $\phi_0^I (t)$, resulting from (\ref{Action_gen}), are the following. The Einstein equations are:
\be \label{EinstEqs}
G_{IJ} \dot{\phi}^I_0 \dot{\phi}^J_0 = - 2 \dot{H}  \qquad  ,  \qquad  3 H^2 + \dot{H} = V \qquad .
\ee
The field equations for the scalars are:
\be \label{EoM_sc}
D_t \dot{\phi}^I_0 + 3 H \dot{\phi}_0^I + G^{IJ} V_J = 0 \quad ,
\ee
where $V_J \equiv \pd_J V$, $\pd_J \equiv \pd_{\phi_0^J}$ and
\be
D_t A^I \equiv \dot{\phi}_0^J \,\nabla_J A^I = \dot{A}^I + \Gamma^I_{JK} \dot{\phi}_0^J A^K
\ee
with $A^I$ being any vector in field space and $\Gamma^I_{JK}$ denoting the Christoffel symbols for the metric $G_{IJ}$. 

We will be interested in the case when the field space, parametrized by $\{\phi^I\}$, is rotationally-invariant. In that case, the metric $G_{IJ}$ can be written as:
\be \label{Gmetric}
ds^2_{G} = d\varphi^2 + f(\varphi) d\theta^2 \,\,\, ,
\ee
where we have introduced the notation:
\be \label{Backgr_id}
\phi^1_0 (t) \equiv \varphi (t) \qquad , \qquad \phi^2_0 (t) \equiv \theta (t)
\ee
and $f(\varphi) \ge 0$ for any $\varphi$\,. With (\ref{Gmetric})-(\ref{Backgr_id}), equations (\ref{EinstEqs}) and (\ref{EoM_sc}) give respectively:
\be \label{EinstEqf}
\dot{\varphi}^2 + f \dot{\theta}^2 = - 2 \dot{H} \qquad , \qquad 3H^2 + \dot{H} = V
\ee
and
\be \label{ScalarEoMs}
\ddot{\varphi} - \frac{f'}{2} \dot{\theta}^2 + 3 H \dot{\varphi} + \pd_{\varphi} V = 0 \qquad , \qquad \ddot{\theta} + \frac{f'}{f} \dot{\varphi} \dot{\theta} + 3 H \dot{\theta} + \frac{1}{f} \pd_{\theta} V = 0  \quad ,
\ee
where we have denoted $f' \equiv \pd_{\varphi} f$\,.

\subsection{Dark energy from large turning in field space} \label{DE_lt}

In multi-field models of cosmic acceleration one can decouple observational constraints, related to the Hubble parameter and its derivatives, from flatness conditions on the scalar potential \cite{AP}. The key for this is in considering non-geodesic background trajectories in field space. 

To describe the deviation from a geodesic, it is convenient to introduce an orthonormal basis of tangent, $T^I$, and normal, $N_I$\,, vectors to a field-space trajectory. In the two-field case, with a background trajectory $(\phi^1_0(t),\phi^2_0(t))$\,, the relevant basis can written as \cite{CAP}:
\be \label{basisTN}
T^I = \frac{\dot{\phi}^I_0}{\dot{\phi}_0} \quad , \quad N_I = (\det G)^{1/2} \epsilon_{IJ} T^J \quad , \quad \dot{\phi}_0^2 = G_{IJ} \dot{\phi}^I_0 \dot{\phi}^J_0 \,\,\, .
\ee
Then one can define the turning rate of the trajectory, which is the quantity measuring its deviation from a geodesic, as \cite{AAGP}:
\be
\Omega = - N_I D_t T^I \,\,\, .
\ee
On solutions of the equations of motion (\ref{EoM_sc}) this is equivalent with:
\be \label{Om_NV}
\Omega = \frac{N_I V^I}{\dot{\phi}_0} \,\,\,\, .
\ee
For a field-space metric of the form (\ref{Gmetric}), the expression (\ref{Om_NV}) gives \cite{LA}:
\be \label{Om_f}
\Omega = \frac{\sqrt{f}}{\left( \dot{\varphi}^2 + f \dot{\theta}^2 \right)} \left[ \dot{\theta} \pd_{\varphi} V - \frac{\dot{\varphi}}{f} \pd_{\theta} V \right] \,\,\, .
\ee

Recently, \cite{ASSV} proposed a construction of dark energy models, relying on strongly non-geodesic motion in field space, even with a steep scalar potential. To show how this proposal works, \cite{ASSV} considered a rotationally-invariant field-space metric and a potential $V (\varphi , \theta)$\,, which is linear in the angular variable $\theta$ and has a minimum at some fixed non-vanishing value of $\varphi$\,. The reason for the non-periodicity in $\theta$ was that taking a potential, compatible with the $U(1)$ invariance of the metric (\ref{Gmetric}), implies a conserved quantity. Namely, one has that \,$a^3 f (\varphi) \dot{\theta} = const$ \,on solutions of the equations of motion. Thus, there are no stable circular trajectories in field space for solutions with an expanding space-time. So the assumption of \cite{ASSV} was that, to realize their proposal, one has to break the periodicity of \,$\theta$ \,inside the potential, in order to obtain circular field-space orbits with \,$\dot{\theta} \approx const$\,. 

In principle, one could justify scalar potentials $V$, which are not well-defined over the entire field-space, by viewing them as belonging to effective descriptions with limited validity. However, in such a case, one would expect $V$ to be well-defined over the part of field space, containing the relevant background trajectory. In the dark energy proposal of \cite{ASSV}, though, one relies on a constantly-large turning rate \,$|\Omega (t)| = |\dot{\theta} (t)| \approx const$\,, and thus on (infinitely) many rotations around the center of field space. Hence, this raises the question whether it is possible to realize such a dark energy model, when the potential $V(\varphi,\theta)$ is well-defined on the surface (\ref{Gmetric}). We will show in the following that the answer is affirmative.\footnote{Our considerations will focus only on (\ref{EinstEqf})-(\ref{ScalarEoMs}), without additional matter fields, since, as explained in \cite{ASSV}, such fields only suppress further the $\varepsilon$-parameter relevant for dark energy.}

In addition to looking for solutions with $\dot{\theta} = const$\,, we will assume that $\pd_{\theta} V = 0$\,. Thus, we will aim to preserve not only the periodicity in $\theta$\,, but also the $U(1)$ isometry of the field space metric (\ref{Gmetric}). At first sight, it might seem that the assumption $\pd_{\theta} V = 0$ would be inconsistent with \,$\dot{\theta} \equiv \omega = const$ \,due to the second equation of (\ref{ScalarEoMs}), which implies:
\be \label{H_intr}
H = - \frac{1}{3} \left( \frac{f'}{f} \dot{\varphi} + \frac{1}{\omega f} \pd_{\theta} V \right) \, .
\ee
So having a finite Hubble parameter $H$ would seem to require $\pd_{\theta} V \neq 0$\,, for a fixed $\varphi$\,, in accordance with the discussion in \cite{ASSV}. However, in principle, it is possible to have solutions with $\dot{\theta} = const$ and $\varphi = \varphi (t)$\,, such that the expression $\frac{f'}{f} \dot{\varphi}$ is always finite and tends to a non-vanishing constant at late times.\footnote{This is similar to the behavior of the solutions investigated in \cite{LA} for the purposes of primordial black hole generation. In that case, one also had $\pd_{\theta} V = 0$\,.} Note that taking $\pd_{\theta} V = 0$ in (\ref{H_intr}) gives $H = - \frac{f' \dot{\varphi}}{3f}$\,, which agrees completely with the Hubble parameter, obtained from (\ref{Hp}) by using the conserved quantity $a^3 f (\varphi) \dot{\theta} = const$\hspace*{0.1cm}, as should be the case for consistency.

Summarizing the above discussion, our goal will be to find background solutions, giving long-lasting accelerated space-time expansion, with the ansatze $\dot{\theta} = const$ and $\pd_{\theta} V = 0$\,. As explained above, the corresponding field-space trajectories cannot be circular. Instead, they would likely represent spirals toward the minimum of the potential. As long as such trajectories are not geodesics on the field space under consideration, they will lead to a large turning rate for any $t$, in accordance with the proposal of \cite{ASSV}.

\section{Inspiraling solution with \,$\dot{\theta} = const$} \label{Sol_Generalities}
\setcounter{equation}{0}

We now turn to looking for solutions of the equations of motion (\ref{EinstEqf})-(\ref{ScalarEoMs}) with the Ansatze $\dot{\theta} = const$ and $\pd_{\theta} V = 0$\,. In this Section we will show that an exact solution of this type does exist for a certain choice of field space metric (\ref{Gmetric}), while deriving in the process the form of the scalar potential that is compatible with it. We will also show that the late-time asymptotic behavior of this solution is precisely what is needed for a dark energy model. 

\subsection{Field equations and scalar potential} \label{SolEx}

Let us now impose the Ansatze:
\be \label{om}
\dot{\theta} \equiv \omega = const \qquad {\rm and} \qquad \pd_{\theta} V = 0
\ee
and analyze their consequences for the background equations of motion (\ref{EinstEqf})-(\ref{ScalarEoMs}).

Substituting (\ref{om}), the second equation in (\ref{ScalarEoMs}) gives:
\be \label{Hf}
H = - \frac{f'}{3f} \dot{\varphi} \,\,\, .
\ee
Using this, together with (\ref{om}), in the first equations of (\ref{ScalarEoMs}) and (\ref{EinstEqf}), we obtain respectively:
\be \label{EoM_ph}
\ddot{\varphi} - \frac{f'}{2} \omega^2 - \frac{f'}{f} \dot{\varphi}^2 + \pd_{\varphi} V = 0 \,\,\, ,
\ee
and
\be \label{EinE1}
\dot{\varphi}^2 + f \omega^2 = \frac{2}{3} \left( \frac{f''}{f} - \frac{f'^2}{f^2} \right) \dot{\varphi}^2 + \frac{2}{3} \frac{f'}{f} \,\ddot{\varphi} \,\,\, .
\ee
Instead of working with the second equation in (\ref{EinstEqf}), it is more convenient to study the following combination of the two equations there:
\be \label{EE}
3 H^2 - \frac{1}{2} \left( \dot{\varphi}^2 + f \dot{\theta}^2 \right) = V \,\,\, .
\ee
Substituting (\ref{om})-(\ref{Hf}), equation (\ref{EE}) acquires the form:
\be \label{E1}
\left( \frac{f'^2}{3f^2} - \frac{1}{2} \right) \dot{\varphi}^2 - \frac{1}{2} f \omega^2 = V \,\,\, .
\ee

Now let us solve (\ref{EoM_ph}) algebraically for $\ddot{\varphi}$ and substitute the result in (\ref{EinE1}). This gives:
\be \label{E2}
\left( 1 - \frac{2 f''}{3 f} \right) \dot{\varphi}^2 = \frac{1}{3} \frac{f'^2}{f} \omega^2 - \frac{2}{3} \frac{f'}{f} \pd_{\varphi} V - f \omega^2 \,\,\, .
\ee
Combining (\ref{E1}) and (\ref{E2}), we obtain:
\be \label{fV}
- \frac{\left( \frac{f''}{3f} - \frac{1}{2} \right)}{\left( \frac{f'^2}{3 f^2} - \frac{1}{2} \right)} = \frac{\frac{1}{6} \frac{f'^2}{f} \omega^2 - \frac{1}{3} \frac{f'}{f} \pd_{\varphi} V - \frac{1}{2} f \omega^2}{V + \frac{1}{2} f \omega^2} \,\,\, .
\ee
This relation can be viewed as an ODE that determines the form of the potential $V(\varphi)$\,, which allows solutions compatible with (\ref{om}) for a given field-space metric (\ref{Gmetric}). Solving analytically (\ref{fV}) for $V (\varphi)$ seems daunting for an arbitrary function $f(\varphi)$\,. However, one can achieve considerable simplification, if one takes $f$ to be such that the metric (\ref{Gmetric}) is hyperbolic, as we will see shortly.\footnote{Note that (\ref{fV}) can be solved with $f (\varphi) = \varphi^2$\,, i.e. with a flat metric (\ref{Gmetric}). However, the resulting scalar potential is negative-definite in the part of field space, containing its minima.} Interestingly, hyperbolic field spaces have attracted a lot of attention in the literature on cosmological inflation; see the original works on $\alpha$-attractors \cite{KLR,KLR2,KL3,CKLR}, as well as the subsequent wide generalizations \cite{LS,BL,BL2}. So it is, perhaps, not surprising to encounter them in the context of dark energy.

In light of the above, we assume from now on that the metric (\ref{Gmetric}) is hyperbolic. Recall that the Gaussian curvature $K$ of a hyperbolic surface is constant and negative by definition. Computing $K$ for (\ref{Gmetric}), we have:
\be
K = - \frac{1}{4} \left( 2 \frac{f''}{f} - \frac{f'^2}{f^2} \right) \,\,\, .
\ee
Therefore, we can rewrite the combination, appearing in the numerator on the left-hand side of (\ref{fV}), as:
\be \label{rel}
\frac{f''}{3f} - \frac{1}{2} = \frac{1}{2} \left[ \frac{1}{3} \frac{f'^2}{f^2} - \frac{4}{3} K - 1 \right] \,\,\, ,
\ee
where $K$ is an arbitrary negative constant. Now, to simplify (\ref{fV}), let us choose:
\be \label{K}
K = - \frac{3}{8} \,\,\, ,
\ee
in which case (\ref{rel}) becomes:
\be
\frac{f''}{3f} - \frac{1}{2} = \frac{1}{2} \left[ \frac{1}{3} \frac{f'^2}{f^2} - \frac{1}{2} \right] \,\,\, .
\ee
Substituting this in (\ref{fV}), we obtain:
\be \label{fV_s}
- \frac{1}{2} V + \frac{1}{4} f \omega^2 = \frac{1}{6} \frac{f'^2}{f} \omega^2 - \frac{1}{3} \frac{f'}{f} \pd_{\varphi} V \,\,\, .
\ee

To solve (\ref{fV_s}) explicitly for $V (\varphi)$, we need first to specify $f (\varphi)$\,. There are three types of rotationally-invariant elementary hyperbolic surfaces: the Poincar\'e disk, the hyperbolic punctured disk and the hyperbolic annulus. We have checked that only the Poincar\'e disk case leads to the desired solution.\footnote{The punctured disk case leads to a Hubble parameter that tends exponentially fast to zero (together with $\varphi (t)$ tending fast to infinity), while in the annulus case there are no background solutions which are periodic in $\theta$.} So let us take the form of $f$\,, which corresponds to the Poincar\'e-disk metric with Gaussian curvature as in (\ref{K}), namely (see, for ex. \cite{ABL}):
\be \label{fs}
f (\varphi) \, = \, \frac{8}{3} \,\sinh^2 \!\left( \sqrt{\frac{3}{8}} \,\varphi \right) \,\,\, .
\ee
Then, (\ref{fV_s}) gives:
\be \label{Vs}
V (\varphi) \,\, = \,\, C_V \,\cosh^2 \!\left( \frac{\sqrt{6}}{4} \,\varphi \right) - \frac{4}{3} \,\omega^2 \quad , \quad C_V = const \,\,\, .
\ee
Note that $V|_{\varphi = 0} = C_V - \frac{4}{3} \,\omega^2$\,. Hence to ensure that the potential is  positive definite, we need to take:
\be \label{CvO}
C_V > \frac{4}{3} \,\omega^2 \ \,.
\ee
We should also point out that, modulo the constant $\omega^2$-term, (\ref{fs}) and (\ref{Vs}) are exactly the same as the respective expressions for $f(\varphi)$ and $V(\varphi)$, required by the hidden symmetry of \cite{ABL}. This will play an important role in the next Section.

Knowing the functions $f$ and $V$ explicitly enables one, in principle, to find $\varphi (t)$ from any of (\ref{EoM_ph}), (\ref{EinE1}) or (\ref{E1}). However, it seems rather challenging to solve analytically any of these three equations, with (\ref{fs}) and (\ref{Vs}) substituted. So, instead of attacking this problem directly, in the next Section we will simplify the equations of motion greatly, by exploiting the close similarity to the hidden symmetry case that we mentioned above. This will enable us to find the analytical form of the solution quite easily.

Even without having $\varphi (t)$ explicitly, though, one can verify that it is not overdetermined by the combination of (\ref{EoM_ph}), (\ref{EinE1}) and (\ref{E1}), thus completing the proof that a solution of the above type exists. Indeed, solving (\ref{E1}) algebraically for $\dot{\varphi}$, one has:
\be \label{ph_d}
\dot{\varphi}^2 = \frac{V + \frac{1}{2} f \omega^2}{\frac{f'^2}{3f^2}-\frac{1}{2}} \,\,\, .
\ee
Substituting (\ref{fs})-(\ref{Vs}) in (\ref{ph_d}), one obtains $\dot{\varphi} = \dot{\varphi} (\varphi)$\,. Taking $\frac{d}{dt}$ of (\ref{ph_d}), we can also find $\ddot{\varphi} = \ddot{\varphi} (\varphi)$\,. Using these $\dot{\varphi} (\varphi)$ and $\ddot{\varphi} (\varphi)$ expressions, as well as (\ref{fs})-(\ref{Vs}), one can easily verify that each of the equations (\ref{EoM_ph}) and (\ref{EinE1}) is automatically satisfied. Thus, indeed, there is a single independent ODE determining $\varphi (t)$\,.

\subsection{Late-time behavior} \label{Late-time}

In the previous subsection, we established the existence of a solution of the equations of motion, satisfying the ansatz (\ref{om}). To realize a two-field dark energy model, in the vein of the type discussed in Section \ref{DE_lt}, this solution should have a dimensionless turning rate $\Omega/H >\!\!> 1$ for a large or unlimited amount of time. Interestingly, we do not need the explicit form of the function $\varphi (t)$\,, in order to show that $\Omega/H$ behaves in this manner. 

To begin with, let us observe that the minimum of the scalar potential (\ref{Vs}) is at $\varphi = 0$\,. Therefore, all solutions compatible with (\ref{Vs}) should tend, with time, to the center of field space. Thus, the large-$t$ limit corresponds to small $\varphi$\,. In addition, (\ref{ph_d}) allows us to rewrite $\Omega (t)$ and $H (t)$ as functions of $\varphi$\,. Indeed, the expression for the turning rate (\ref{Om_f}), with (\ref{om}) substituted, is:
\be \label{Om}
\Omega \,\, = \,\frac{f^{1/2}(\varphi) \,\, \omega \,\, \pd_{\varphi} V (\varphi)}{\dot{\varphi}^2 \,+ \,f(\varphi) \,\omega^2} \,\,\, .
\ee
Substituting (\ref{ph_d}), together with (\ref{fs})-(\ref{Vs}), into (\ref{Hf}) and (\ref{Om}), we obtain $H = H (\varphi)$ and $\Omega = \Omega (\varphi)$, respectively. Taking the limit $\varphi \rightarrow 0$ in these functions, we find:
\be \label{Hlt}
H^2 \, \rightarrow \, \frac{1}{3} C_V - \frac{4}{9} \omega^2 + {\cal O} (\varphi^2) 
\ee
and
\be
\Omega^2 \, \rightarrow \, \omega^2 + {\cal O} (\varphi^2) \,\,\, .
\ee
Hence, at late times we have:
\be
\left( \frac{\Omega}{H} \right)^2 \, \rightarrow \, \frac{9 \omega^2}{3 C_V - 4 \omega^2} + {\cal O} (\varphi^2) \,\,\, .
\ee
Clearly, one can obtain as large a dimensionless turning rate as desired by choosing suitably the arbitrary constants $C_V$ and $\omega$, in a manner consistent with (\ref{CvO}). 

We can also compute the large $t$ behavior of the slow roll parameter $\varepsilon = - \dot{H}/H^2$\,. Indeed, differentiating (\ref{Hf}) with respect to $t$ gives:
\be \label{Hd}
\dot{H} = - \frac{1}{3} \left( \frac{f''}{f} - \frac{f'^2}{f^2} \right) \dot{\varphi}^2 - \frac{1}{3} \frac{f'}{f} \ddot{\varphi} \,\,\, .
\ee
Using again the expressions $\dot{\varphi} (\varphi)$ and $\ddot{\varphi} (\varphi)$ that follow from (\ref{ph_d}), we can rewrite $\dot{H} (t)$ in (\ref{Hd}) as a function of $\varphi$\,. Taking the limit $\varphi \rightarrow 0$ in that function, we obtain:
\be
\dot{H} \, \rightarrow \, - \frac{3 C_V}{8} \varphi^2 + {\cal O} (\varphi^4) \,\,\, .
\ee
Therefore, at large $t$ (equivalently, small $\varphi$) we have:
\be \label{Ep_late_t}
\varepsilon = - \frac{\dot{H}}{H^2} \,\, \rightarrow \,\, \frac{9 C_V}{8(C_V -\frac{4}{3}\omega^2)} \,\varphi^2 + {\cal O} (\varphi^4) \,\,\, ,
\ee
implying that $\varepsilon$ tends to zero. Hence, at late times the slow roll condition $\varepsilon <\!\!< 1$ is very well satisfied. And furthermore, it is improving with time.

To recapitulate, we have shown that the solution of the previous subsection has to behave as desired for a dark energy model. Let us now turn to finding it analytically.

\section{Finding the exact inspiraling solution} \label{ExactSol}
\setcounter{equation}{0}

The considerations of Section \ref{Sol_Generalities} led us to choose $f(\varphi)$ such that the field-space metric (\ref{Gmetric}) is that of the Poincar\'e disk with Gaussian curvature fixed as in (\ref{K}), precisely as required by the hidden symmetry of \cite{ABL}. The resulting scalar potential (\ref{Vs}) breaks that hidden symmetry only due to the constant $\omega^2$ term in it, as noted above. This close similarity to the hidden symmetry case will allow us to find the exact solution analytically by applying techniques from the Noether symmetry method.

To begin with, let us substitute the background ansatze (\ref{metric_g}) and the field-space metric (\ref{Gmetric}) in the action (\ref{Action_gen}). After an integration by parts, we obtain the Lagrangian density:
\be \label{L_class_mech}
{\cal L} \,= \,- 3 a \dot{a}^2 + \frac{a^3 \dot{\varphi}^2}{2} + \frac{a^3 f (\varphi) \,\dot{\theta}^2}{2} - a^3 V(\varphi,\theta)
\ee
per unit spatial volume, which can be viewed as a classical mechanical Lagrangian for the generalized coordinates $a$, $\varphi$ and $\theta$\,. Let us now make the same change of variables, on the space of generalized coordinates, as in the Poincar\'e disk case of \cite{ABL} (see also the concise summary in Section 4 of \cite{LA}), namely:\footnote{We will use the slightly more convenient notation introduced in \cite{LA}.}
\bea \label{Ch_var}
a (t) &=& \left[ u^2 - \left( v^2 + w^2 \right) \right]^{1/3} \,\,\,\, , \nn \\
\varphi (t) &=& \sqrt{\frac{8}{3}} \,{\rm arccoth} \!\left( \sqrt{\frac{u^2}{v^2 + w^2}} \,\,\right) \,\,\,\, , \nn \\
\theta (t) &=& \theta_0 + {\rm arccot} \!\left( \frac{v}{w} \right) \,\,\,\, , \,\,\,\, \theta_0 = const \quad .
\eea
Substituting (\ref{Ch_var}), together with (\ref{fs}) and (\ref{Vs}), inside (\ref{L_class_mech}), we find:
\be \label{Lagr_uvw}
{\cal L} \,= \,- \frac{4}{3} \dot{u}^2 + \frac{4}{3} \dot{v}^2 + \frac{4}{3} \dot{w}^2 - \frac{4}{3} \kappa^2 u^2 - \frac{4}{3} \omega^2 v^2 - \frac{4}{3} \omega^2 w^2 \,\,\,\, ,
\ee
where
\be
\kappa \equiv \frac{1}{2} \sqrt{3C_V - 4 \omega^2} \,\,\, .
\ee
As was to be expected, in (\ref{Lagr_uvw}) there is no cyclic variable, since the hidden symmetry of \cite{ABL} is broken by the constant term in our potential (\ref{Vs}). Nevertheless, (\ref{Ch_var}) has led to a great simplification.

The Euler-Lagrange equations of (\ref{Lagr_uvw}) have the following general solutions:
\bea \label{Sols_uvw}
u (t) &=& C_1^u \sinh (\kappa t) + C_0^u \cosh (\kappa t) \quad , \nn \\
v (t) &=& C_1^v \sin (\omega t) + C_0^v \cos(\omega t) \quad , \nn \\
w (t) &=& C_1^w \sin (\omega t) + C_0^w \cos(\omega t) \quad ,
\eea
where $C_{0,1}^{u,v,w}$ are integration constants. Recall, however, that we are looking for a solution with $\dot{\theta} = const$\,. From (\ref{Ch_var}), we have:
\be \label{Th_d}
\dot{\theta} = \frac{v \dot{w} - w \dot{v}}{v^2 + w^2} \,\,\,\, .
\ee
To ensure that (\ref{Th_d}) gives $\dot{\theta} = \omega$\,, or equivalently that $\theta - \theta_0$ in (\ref{Ch_var}) is linear in $t$\,, we take in (\ref{Sols_uvw}):
\be \label{CCh}
C_1^v = 0 \quad , \quad C_0^w = 0 \quad {\rm and} \quad C_0^v = C_1^w \equiv C_w \quad ,
\ee
thus obtaining:
\be \label{Sols_vw}
v (t) = C_w \cos(\omega t) \quad \,, \,\quad w (t) = C_w \sin(\omega t) \quad .
\ee
Clearly, substituting (\ref{Sols_vw}) inside (\ref{Ch_var}), we have:
\be
\theta (t) = \theta_0 + \omega t \,\,\,\, ,
\ee
as desired. 

The expressions in (\ref{Ch_var}), with $u(t)$, $v(t)$ and $w(t)$ as above, solve the scalar field equations (\ref{ScalarEoMs}). As is well known, to ensure that the Einstein equations (\ref{EinstEqf}) are also satisfied, one has to supplement the Euler-Lagrange equations of (\ref{Lagr_uvw}) with the Hamiltonian constraint:
\be \label{Constr}
E_{\cal L} = 0 \qquad , \qquad E_{\cal L} \equiv \frac{\pd {\cal L}}{\pd \dot{u}} \dot{u} + \frac{\pd {\cal L}}{\pd \dot{v}} \dot{v} + \frac{\pd {\cal L}}{\pd \dot{w}} \dot{w} - {\cal L} \quad .
\ee
Substituting $u$ from (\ref{Sols_uvw}) and $v,w$ from (\ref{Sols_vw}), we find that (\ref{Constr}) acquires the form: 
\be \label{Constr_c}
\kappa^2 \!\left[ (C_0^u)^2 - (C_1^u)^2 \right] + 2 \omega^2 C_w^2 = 0 \,\,\, .
\ee
Clearly, to be able to satisfy this condition, one has to take:
\be \label{C_C1_C0_u}
|C_1^u| > |C_0^u| \,\,\, .
\ee

It is important to note that the choices in (\ref{CCh}) do not represent any loss of generality in our case. Indeed, the number of independent constants we are left with in this Section, when taking into account (\ref{Constr_c}), is exactly the same as the number of independent constants that the solutions of Section \ref{Sol_Generalities} should have.

Finally, as a consistency check, it is easy to verify that, with the functions $\varphi (t)$ and $a(t)$ obtained in this Section, (\ref{Hf}) gives exactly the same expression for the Hubble parameter as (\ref{Hp}), as should be the case.

\section{Parameter space of the solution} \label{ParSpace}
\setcounter{equation}{0}

In the previous Section we found a new class of exact solutions to the equations of motion (\ref{EinstEqf})-(\ref{ScalarEoMs}), with $f$ and $V$ given by (\ref{fs}) and (\ref{Vs}) respectively. Let us summarize these results. Collecting (\ref{Ch_var}), (\ref{Sols_uvw}), (\ref{CCh}) and (\ref{Constr_c}), we have the solutions:
\bea \label{Sol_om}
a (t) &=& \left( u^2 - C_w^2 \right)^{1/3} \,\,\,\, , \nn \\
\varphi (t) &=& \sqrt{\frac{8}{3}} \,{\rm arccoth} \!\left( \sqrt{\frac{u^2}{C_w^2}} \,\,\right) \,\,\,\, , \nn \\
\theta (t) &=& \theta_0 + \omega t \,\,\,\, ,
\eea
where 
\be \label{Sol_u_om}
u (t) = C_1^u \sinh (\kappa t) + C_0^u \cosh (\kappa t) \qquad {\rm with} \qquad \kappa \equiv \frac{1}{2} \sqrt{3C_V - 4 \omega^2}
\ee
and the integration constants satisfy the relation:
\be \label{Constr_cp}
\kappa^2 \!\left[ (C_0^u)^2 - (C_1^u)^2 \right] + 2 \omega^2 C_w^2 = 0 \,\,\,\, .
\ee
Note that, since an overall numerical factor in $a (t)$ cancels inside $H (t) = \dot{a} / a$\,\,, one can set $C_w = 1$ in (\ref{Sol_om})-(\ref{Constr_cp}) without affecting the physics of the problem.\footnote{Rescaling the arbitrary constants $C_{0,1}^u$ as follows: $C_{0,1}^u \rightarrow C_w C_{0,1}^u$\,, we can cancel $C_w$ in $\varphi (t)$ and in the constraint (\ref{Constr_cp}), while obtaining $a(t) \rightarrow (C_w)^{2/3} (u^2 - 1)^{1/3}$\,.} However, in the following, it will turn out to be technically useful to keep the constant $C_w$ arbitrary.

Let us now discuss additional physical constraints on the parameter space of the solution (\ref{Sol_om})-(\ref{Constr_cp}), as well as investigate their implications for the corresponding field-space trajectories and $\varepsilon$-parameter.

\subsection{Physical parameter space}

In order to ensure the physical conditions that $a(t)$ and $\dot{a}(t)$ are real and positive for any $t \ge 0$\,, where $t=0$ is the initial moment of the expansion regime under consideration\footnote{We take $t=0$ to be the time when the dark energy sector starts driving the expansion of the Universe. At earlier times matter, and before that radiation etc., dominates the evolution and the fields $\varphi$, $\theta$ are effectively frozen due to the large Hubble friction; see, for instance, \cite{ASSV}.}, we need to impose further constraints on the parameter space of the above solutions. To understand that, let us compute from (\ref{Sol_om}):
\be
a(t)|_{t=0} = \left[ (C_0^u)^2 - C_w^2 \right]^{1/3} \quad , \,\quad \dot{a}(t)|_{t=0} = \frac{2}{3} \frac{\kappa C_0^u C_1^u}{\left[ (C_0^u)^2 - C_w^2 \right]^{2/3}} \quad .
\ee
Hence, to have $a(t) > 0$ and $\dot{a}(t) > 0$ for $\forall t \in [0, \infty)$\,, we need to take:
\be \label{Phys_c}
|C_0^u| > |C_w| \qquad {\rm and} \qquad C_0^u C_1^u > 0 \quad .
\ee
Note that the second condition in (\ref{Phys_c}) implies, in particular, that $C_0^u$ and $C_1^u$ have the same sign. To see that the constraints (\ref{Phys_c}) indeed guarantee positive and monotonically increasing $a(t)$, let us consider the function $a^3 (t) = u^2 (t) - C_w^2$ with $u(t)$ as in (\ref{Sol_u_om}). Due to the second condition in (\ref{Phys_c}), we have $\frac{d}{dt} (u^2 - C_w^2) = 2 u \dot{u} > 0$ \,for $\forall t\ge 0$\,. In addition, using the first condition in (\ref{Phys_c}), we conclude that $(u^2 - C_w^2)|_{t=0}=(C_0^u)^2 - C_w^2 \,> \,0$\,. Therefore, $(u^2 - C_w^2) > 0$ for $\forall t \in [0,\infty)$\,, everywhere in the physical parameter space. Thus, we have shown that both $a$ and $\dot{a}$ are positive for any $t \ge 0$\,, within the above constraints on the integration constants. Summarizing all restrictions so far, the physical part of the parameter space of the solutions (\ref{Sol_om})-(\ref{Constr_cp}) is determined by (\ref{CvO}), (\ref{C_C1_C0_u}) and (\ref{Phys_c}). 

\begin{figure}[t]
\begin{center}
\includegraphics[scale=0.35]{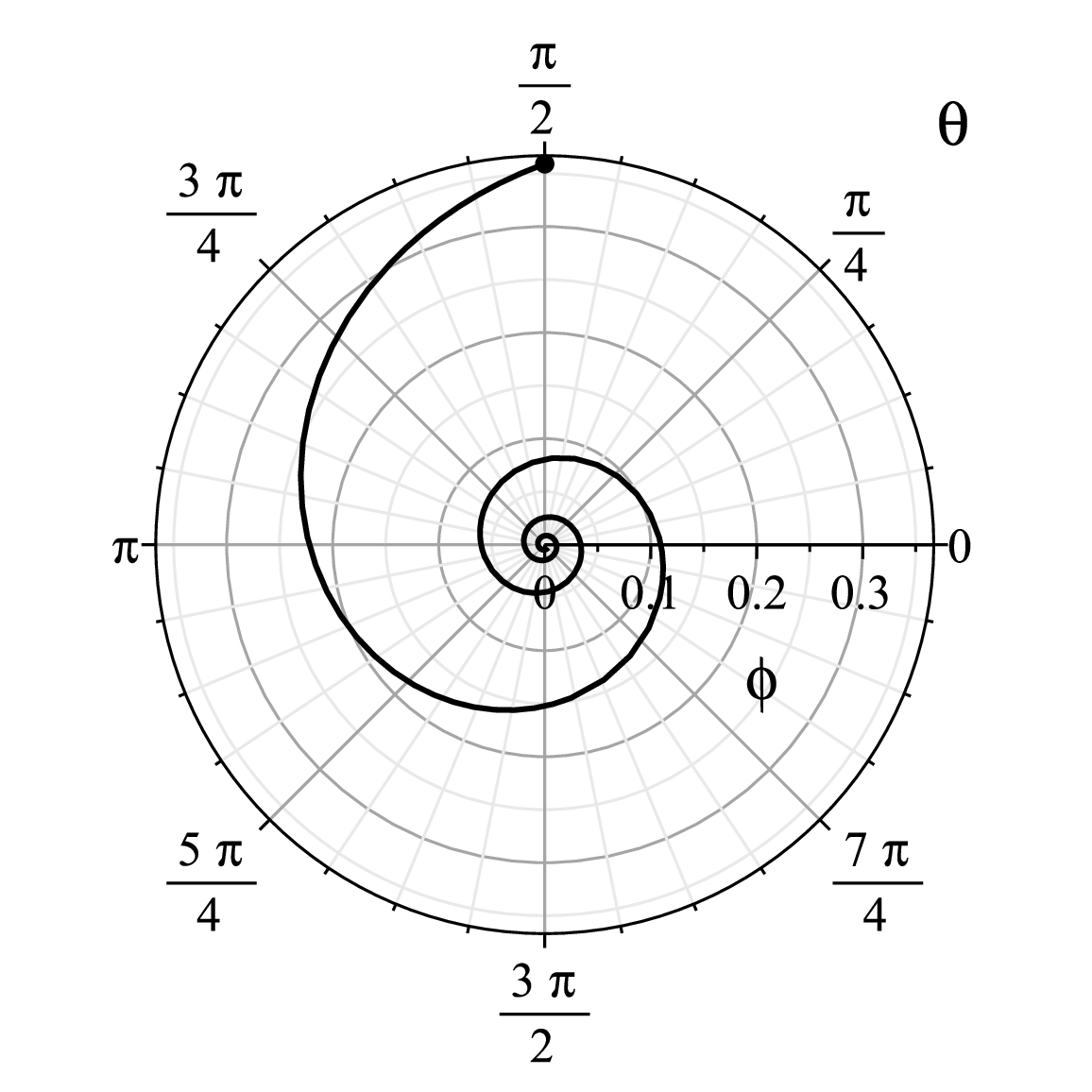}
\hspace*{0.2cm}
\includegraphics[scale=0.35]{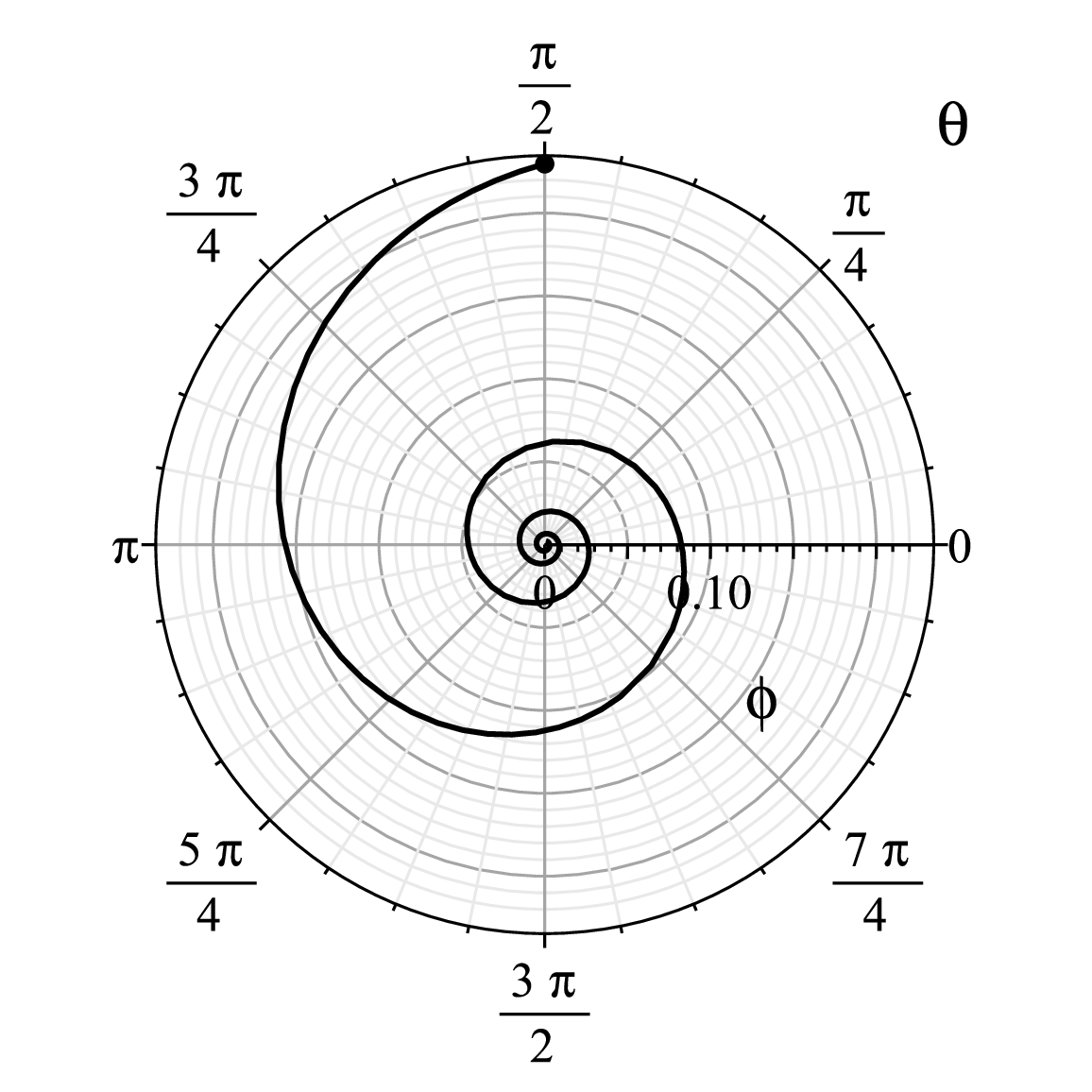}
\end{center}
\vspace{-0.7cm}
\caption{{\small Two examples of trajectories $\left(\varphi(t), \theta(t)\right)$ of the exact solutions (\ref{Sol_om})-(\ref{Constr_cp}), obtained for two sets of parameter choices given in the text. The dot, situated at $\left( \varphi , \theta \right) = \left( 0.36 , \frac{\pi}{2} \right)$ on the left-side plot and at $\left( \varphi , \theta \right) = \left( 0.23 , \frac{\pi}{2} \right)$ on the right-side one, denotes the starting point of the respective trajectory at $t=0$\,.}}
\label{Traj}
\vspace{0.1cm}
\end{figure}
Let us now illustrate the shape of the field-space trajectories $\left(\varphi(t), \theta(t)\right)$ of our solutions. Clearly, there are infinitely many examples consistent with all of the physical constraints above. For convenience, on Figure \ref{Traj} we plot two examples obtained for the following choices of the constants: $C_0^u = 1$, $C_1^u = 2$ (on the left side) and $C_0^u = 2$, $C_1^u = 3$ (on the right side), while the remaining constants in both cases are $\omega = 4$, $C_V = 22$ and $C_w$ is determined by the positive root of (\ref{Constr_cp}). As already anticipated, the trajectories represent spiraling inward, toward the center of field-space at $\varphi = 0$\,.

It is easy to see that every trajectory has a similar shape, for any $t\ge 0$\,, everywhere in the physical parameter space of our solutions. Indeed, from (\ref{Sol_om}), we have:
\be \label{phi_u}
\dot{\varphi} (t) \, = \, - \,\frac{2 \sqrt{6}}{3} \,\bigg|\frac{C_w}{u}\bigg| \,\frac{u \,\dot{u}}{\left( u^2 - C_w^2\right)} \,\,\, ,
\ee
where $u$ is given by (\ref{Sol_u_om}). As explained above, the conditions (\ref{Phys_c}) imply that both $u \dot{u} > 0$ and $(u^2 - C_w^2) > 0$ \,for any $t\ge 0$\,. Therefore:
\be \label{phi_d}
\dot{\varphi} (t) < 0
\ee
for any $t \in [0, \infty)$ and for any values of the integration constants, satisfying the physical constraints. Of course, this is consistent with the expectation that at late times $\varphi (t)$ has to tend to zero, since $\varphi = 0$ is the minimum of the scalar potential (\ref{Vs}). However, the result (\ref{phi_d}) is much stronger. Together with $\dot{\theta} (t) = const$\,, it implies that the field-space trajectories of our solutions are spirals toward the center of the Poincar\'e disk, for any values of the integration constants within the physical parameter space.

Finally, note that, at large $t$\,, (\ref{Sol_u_om}) becomes $u (t) \sim e^{\kappa t}$\,. Thus, one can immediately see from (\ref{phi_u}) that $\dot{\varphi} (t) \rightarrow 0$ as $t \rightarrow \infty$\,. Nevertheless, one can easily verify that substituting our solution in (\ref{Hf}) gives exactly the same {\it finite} large-$t$ limit for $H$ as in (\ref{Hlt}). In addition, with $\varphi$ and $\dot{\varphi}$ tending to zero and $V$ as in (\ref{Vs}), it is clear that the equation of state of our solution approaches
\be \label{wDE}
w_{DE} = - 1 \,\,\, ,
\ee
becoming arbitrarily close to (\ref{wDE}) with time; here \,$w_{DE} \equiv \frac{\frac{1}{2} G_{IJ} \dot{\phi}^I_0 \dot{\phi}^J_0-V}{\frac{1}{2}G_{IJ}\dot{\phi}^I_0 \dot{\phi}^J_0+V}$ \,as usual. This is in accord with current observational constraints.

\subsection{Slow roll expansion}

Let us now consider the implications of (\ref{C_C1_C0_u}) and (\ref{Phys_c}) for the behavior of the slow roll parameter $\varepsilon (t) = - \dot{H} / H^2$\,. We will prove that, within the entire physical parameter space of our exact solutions, $\varepsilon (t)$ is a monotonically decreasing function for any $t \ge 0$\,. In other words, we will show that the slow roll approximation continuously improves with time forever.\footnote{Recall that $\varepsilon (t) > 0$ always for any solution of (\ref{EinstEqf}).} This is consistent with the $t \rightarrow \infty$ limit in (\ref{Ep_late_t}). However, now we can prove analytically that $\dot{\varepsilon} (t) < 0$ for $\forall t\in [0,\infty)$\,, everywhere in physical parameter space.

For that purpose, let us compute the expression that the solution (\ref{Sol_om})-(\ref{Sol_u_om}) implies for $\varepsilon (t)$\,. To keep the calculations manageable, we begin by substituting only (\ref{Sol_om}). This gives:
\be
\varepsilon (t) = \frac{3}{2} \,\frac{u^2 \dot{u}^2 - u^3 \ddot{u} + (\dot{u}^2 + u \ddot{u}) C_w^2 }{u^2 \dot{u}^2} \quad .
\ee
Hence:
\be \label{ep_u}
\dot{\varepsilon} (t) = - \frac{3}{2} \,\frac{\left( \dot{u}^2 \ddot{u} - 2 u \ddot{u}^2 + u \dot{u} u^{(3)} \right)}{\dot{u}^3} - \frac{3}{2} \,\frac{\left( 2 \dot{u}^4 - u^2 \dot{u} u^{(3)} + u \dot{u}^2 \ddot{u} + 2 u^2 \ddot{u}^2 \right)}{u^3 \dot{u}^3} \,C_w^2 \quad .
\ee
Now, in view of the discussion below (\ref{Constr_cp}), physical properties should not depend on the value of $C_w$\,. Therefore, if $\dot{\varepsilon} (t)$ is to be negative-definite, then each of the two terms in (\ref{ep_u}) should be negative-definite by itself. So let us consider them separately.

For convenience, we denote the two terms in (\ref{ep_u}) by:
\bea
\dot{\varepsilon}_{T1} \!&\equiv & \!- \,\frac{3}{2} \,\frac{\left( \dot{u}^2 \ddot{u} - 2 u \ddot{u}^2 + u \dot{u} u^{(3)} \right)}{\dot{u}^3} \,\,\,\, , \nn \\
\dot{\varepsilon}_{T2} \!&\equiv & \!- \,\frac{3}{2} \,\frac{\left( 2 \dot{u}^4 - u^2 \dot{u} u^{(3)} + u \dot{u}^2 \ddot{u} + 2 u^2 \ddot{u}^2 \right)}{u^3 \dot{u}^3} \,C_w^2 \,\,\,\, .
\eea
Substituting (\ref{Sol_u_om}) in $\dot{\varepsilon}_{T1}$\,, we find:
\be
\dot{\varepsilon}_{T1} \, = \, - \,\frac{3 \,\kappa \left[(C_1^u)^2 - (C_0^u)^2 \right] \left[ C_1^u \sinh (\kappa t) + C_0^u \cosh (\kappa t) \right]}{\left[ C_1^u \cosh (\kappa t) + C_0^u \sinh (\kappa t) \right]^3} \,\,\,\, .
\ee
Clearly, the sign of this expression is negative for $\forall t \in [0,\infty)$\,, due to (\ref{C_C1_C0_u}) and (\ref{Phys_c}).\footnote{We mean the second condition in (\ref{Phys_c}), which implies that $C_0^u$ and $C_1^u$ have the same sign.} Now, using (\ref{Sol_u_om}) in $\dot{\varepsilon}_{T2}$\,, after some manipulation, gives:
\be \label{ep_T2}
\dot{\varepsilon}_{T2} \, = \, - \,\frac{3 \,\kappa \,\varepsilon_n}{\left[ C_1^u \sinh (\kappa t) + C_0^u \cosh (\kappa t) \right]^3 \left[ C_1^u \cosh (\kappa t) + C_0^u \sinh (\kappa t) \right]^3} \,C_w^2 \,\,\,\, ,
\ee
where
\bea \label{ep_n}
\varepsilon_n \!&=& \!(C_1^u)^4 + (C_0^u)^4 + 4 C_1^u C_0^u \!\left[ (C_1^u)^2 + (C_0^u)^2 \right] \!\left[ \cosh^2 (\kappa t) + \sinh^2 (\kappa t) \right] \!\sinh (\kappa t) \cosh (\kappa t) \nn \\
\!&+& \!2 \!\left[ 6 (C_1^u)^2 (C_0^u)^2 + (C_0^u)^4 + (C_1^u)^4 \right] \!\cosh^2 (\kappa t) \sinh^2 (\kappa t) \,\,\,\, .
\eea
Notice that, due to (\ref{Phys_c}), every term in $\varepsilon_n$ is positive-definite for any $t \ge 0$\,. Therefore, taking into account that $C_0^u$ and $C_1^u$ have the same sign, we conclude that $\dot{\varepsilon}_{T2}$ is negative-definite for $\forall t \in [0,\infty)$\,. 

Summarizing the above considerations, we have shown that $\dot{\varepsilon} (t) < 0$ for any $t\ge 0$\,, everywhere in physical parameter space. Note that keeping the constant $C_w$ arbitrary provided a useful technical guidance on how to simplify the computations, by splitting the seemingly unwieldy expression for $\dot{\varepsilon} (t)$ into two more easily manageable parts.

Notice that the scale factor of the solutions (\ref{Sol_om})-(\ref{Sol_u_om}) approaches fast the late-time behavior $a(t) \sim \exp \!\left( \frac{2}{3}\kappa t \right)$\,. The result of this Section, that $\varepsilon (t)$ is a monotonically decreasing function, means that our solutions tend to this de Sitter expansion continuously with time forever. The lack of a natural exit from the accelerated expansion phase would be a problem for an inflationary model. However, it is a desirable feature for a model of dark energy.

\section{Discussion} \label{Discussion}

We found a class of exact solutions to the cosmological background equations of motion, which realize a model of two-field dark energy. These solutions rely on a Poincar\'e-disk scalar manifold and a rotationally-invariant scalar potential. Although their trajectories in field space represent spiraling inward, toward $\varphi = 0$\,, the corresponding Hubble parameter is always finite and tends fast to a constant.\footnote{Its magnitude can be varied at will by choosing suitably certain integration constants.} As discussed in \cite{ASSV}, at the level of the classical background, there is little difference between such a model of cosmic acceleration and a pure cosmological constant. However, at the level of perturbations around that background, the new model can lead to important deviations from $\Lambda$CDM. Namely, a reduced sound speed of scalar perturbations can cause clustering of dark energy on sub-horizon scales, affecting large-scale structure formation. It would, undoubtedly, be of great interest to investigate in depth the cosmological consequences of this model, as well as how they compare to previous studies \cite{HS,MT,CDNSV,HRCMAL,HAK} of clustering dark energy.

Although we were motivated by interest in the late-time Universe, our results may also have relevance for cosmological inflation. Indeed, they show that it is possible to obtain long-term accelerated expansion from strongly non-geodesic field space trajectories, which are not circular. This point is conceptually important regarding the issue of embedding inflationary models, which rely on large turning rates, into supergravity. Such an embedding would be a first step toward understanding how to construct rapid-turning models in string compactifications. Since quantum-gravitational arguments provide motivation for multi-field inflation, embeddability in supergravity may be a criterion for whether a certain model belongs to the landscape or swampland of effective theories.\footnote{It is far from clear, though, that supergravity would be directly relevant for the late-time Universe and thus for models of dark energy, although see \cite{BQ}.} 
It was suggested recently that embedding rapid-turn inflation in supergravity is rather difficult,
because it requires a negative eigenvalue of the classical mass matrix, as well as large field-space curvature \cite{ACPRZ}. A starting assumption of these considerations was that $\dot{\varphi} = 0$\,.\footnote{Actually, this is one of a number of simplifying assumptions in \cite{ACPRZ}.} The reason is that the typical large-turning inflationary (approximate) solutions in the literature have field-space trajectories with radii, which are (almost) fixed at large values; see for instance \cite{ARB,CRS,AW}. However, in our exact solutions, which can have as large a turning rate as desired, one should never substitute $\dot{\varphi} = 0$\,, despite $\dot{\varphi}$ tending fast to zero with time. Indeed, as we have seen, $\varphi$ tends to zero as well, and in such a manner that the entire combination in (\ref{Hf}) always gives a finite Hubble parameter.\footnote{This is similar to the behavior of the inflationary solutions investigated in \cite{LA} for the purposes of PBH-generation, although in that case the turning rates are large only briefly. In those solutions, the Hubble parameter is also always finite, while both $\varphi$ and $\dot{\varphi}$ ultimately tend to zero. Unlike in \cite{LA}, though, in the present case the large turning rate does not lead to a tachyonic instability of the entropic perturbation. Indeed, one can easily verify that the effective entropic mass $m_s^2 (t) = N^I N^J V_{;IJ} - \Omega^2$ is positive, by performing the same kind of considerations as in Section \ref{Late-time}. More precisely, by using (\ref{basisTN}), (\ref{fs}), (\ref{Vs}) and (\ref{ph_d}), one finds that $N^I N^J V_{;IJ}$ tends to $\frac{3}{4} C_V$ from above, while $\Omega^2$ tends to $\omega^2$ from below. Together with (\ref{CvO}), this implies that $m_s^2 > 0$\, always.} This raises a whole new possibility to explore, regarding the question under what conditions supergravity models of inflation can accommodate large turning rates. We hope to address this problem systematically in the future.

It is worth underlining that, unlike in the typical hyperbolic models, in our solutions the curvature of the scalar manifold is not a free parameter. Instead, it is fixed according to (\ref{K}). Interestingly, this is exactly what is required for the presence of a hidden symmetry in two-field models with a scalar metric of the form (\ref{Gmetric}), as shown in \cite{ABL,ABL2}. This is reminiscent of a point of enhanced symmetry in a larger moduli space, which may be suggestive regarding possible embeddings in a more fundamental theoretical framework. In our case, of course, the hidden symmetry of \cite{ABL} is mildly broken by the presence of a constant term in the potential. Nevertheless, it is an interesting open question whether the closeness to this symmetry is coincidental, although technically very useful as we saw in Section \ref{ExactSol}, or it has a deeper meaning. To achieve progress on this issue, it may be beneficial to explore the possibility for inspiraling solutions, whose existence is enabled by (mild breaking of) more general hidden symmetries, such as those of \cite{ABL2}. 

It is also interesting to reconsider the conclusions of \cite{BCHCSJY}, regarding Hubble tension, in the context of multi-field dark energy. Of course, this problem could be resolved, alternatively, by new physics in the Early Universe; see for instance \cite{PSB}, as well as the discussion in \cite{CDP2}. However, our class of dark energy models is rather different from the one underlying the considerations of \cite{BCHCSJY}. So it merits a separate investigation, in order to establish whether the Hubble tension is exacerbated or alleviated in our case. Interestingly, such a study could point to observationally preferable parts of the parameter space of this type of models.

Finally, it is important to note that for our solutions the phenomenologically-relevant part of field space is a neighborhood of $\varphi = 0$\,, unlike for the standard hyperbolic models in the literature, which rely on (very) large field values. Clearly, this makes our models much more appealing theoretically, since they are more reliable as effective descriptions. Of course, whether they are also attractive phenomenologically (i.e., as models of clustering dark energy) is a subject for further detailed studies. We hope to report on this in the near future.

\section*{Acknowledgements}

We would like to thank E. M. Babalic, C. Lazaroiu, S. Paban, P. Suranyi and I. Zavala for interesting discussions. L.A. also thanks the Stony Brook Simons Workshop in Mathematics and Physics for hospitality during the completion of this work. L.C.R.W. would like to thank the Aspen Center for Physics, which is supported by the U.S. NSF grant PHY-1607611, where he benefited from attending the summer program on the physics of the dark sector and having discussions with the participants. L.A. has received partial support from the Bulgarian NSF grants DN 08/3 and KP-06-N38/11. J.D. is supported by a Univ. of Cincinnati scholarship.

\end{document}